\documentclass[11pt,english]{article}
\usepackage{courier}
\usepackage[T1]{fontenc}
\usepackage[latin9]{inputenc}
\usepackage{verbatim}
\usepackage{amsmath}
\usepackage{amssymb}

\makeatletter
\usepackage{hyperref}
\usepackage{amsfonts}
\usepackage{amsthm}
\usepackage{amsfonts}
\usepackage{multirow}
\usepackage{courier}
\usepackage{fullpage}
\usepackage{epsfig}
\usepackage{bm}
\usepackage{times}
\@ifundefined{definecolor}
 {\usepackage{color}}

\newtheorem{theorem}{Theorem}
\newtheorem{lem}{Lemma}
\newtheorem{lemma}[lem]{Lemma}

\newtheorem{obs}{Observation}
\newtheorem{prop}{Proposition}

\theoremstyle{definition}
\newtheorem{defn}{Definition}

\theoremstyle{remark}

\makeatletter\def\imod#1{\allowbreak\mkern10mu({\operator@font mod}\,\,#1)}\makeatother

\newcommand{\eps}{\epsilon}
\newcommand{\cube}{\operatorname{\{0, 1\}}}
\newcommand{\ncube}{\F_{2}^{n}}
\newcommand{\kcube}{\F_{2}^{k}}
\newcommand{\Had}{\text{Had}}
\newcommand{\F}{\mathbb{F}}

\newcommand{\N}{{\mathbb N}}
\newcommand{\calf}{{\cal F}}

\newcommand{\eqdef}{{\stackrel{\rm def}{=}}}

\newcommand{\proP}{\mathcal{P}}
\newcommand{\proPo}{\mathcal{P}_{1}}
\newcommand{\proPt}{\mathcal{P}_{2}}
\newcommand{\ozz}{\mathcal{P}_{\text{$(100)$-$\mathbf{FREE}$}}}
\newcommand{\ooo}{\mathcal{P}_{\text{$(111)$-$\mathbf{FREE}$}}}
\newcommand{\ooz}{\mathcal{P}_{\text{$(110)$-$\mathbf{FREE}$}}}
\newcommand{\lin}{\mathcal{P}_{\mathbf{LIN}}}
\newcommand{\LBP}{\mathcal{P}_{\mathbf{NLTF}}}
\newcommand{\dist}{\text{dist}}
\newcommand{\supp}{\text{supp}}
\newcommand{\poly}{\text{poly}}

\begin{document}

\title{Property Testing via Set-Theoretic Operations}

\author{Victor Chen%
\thanks{Princeton University. \texttt{vychen@princeton.edu}.%
} \and Madhu Sudan%
\thanks{Microsoft Research New England. \texttt{madhu@microsoft.com}.
Research supported in part by NSF Award CCR-0514915.%
} \and Ning Xie%
\thanks{MIT CSAIL. \texttt{ningxie@csail.mit.edu}. Research supported in part
by NSF Award CCR-0514771, 0728645 and 0732334.%
} }

\date{}
\maketitle
\setcounter{page}{0}
\begin{abstract}
Given two testable properties $\mathcal{P}_{1}$ and $\mathcal{P}_{2}$,
under what conditions are the union, intersection or set-difference
of these two properties also testable?
We initiate a systematic study of these basic set-theoretic operations in the context of property
testing.
As an application, we give a conceptually different proof
that linearity is testable, albeit with much worse query complexity.
Furthermore, for the problem of testing disjunction of linear functions,
which was previously known to be one-sided testable with a super-polynomial query
complexity, we give an improved analysis and show it has query complexity $O(1/\eps^2)$, where $\eps$ is the distance parameter.
\end{abstract}

\newpage

\section{\texorpdfstring{Introduction}{1. Introduction}}
\label{Sec:Introduction} 

During the last two decades, the size of data sets has been increasing at an exponential rate, 
rendering a linear scan of the whole input an unaffordable luxury. 
Thus, we need \emph{sublinear time} algorithms that read a vanishingly small 
fraction of their input and still output something intelligent and non-trivial
 about the properties of the input. 
The model of property testing~\cite{RS96,GGR98} has been
very useful in understanding the power of sublinear time. 
Property testing is concerned with the existence of a sublinear time 
algorithm that queries an input object a small number of times and
decides correctly with high probability whether the object has a given
property or whether it is  ``far away'' from having the property.

We model input objects as strings of arbitrary length, 
which can also be viewed as a function on arbitrarily large domain. 
Formally, let $\mathcal{R}$ be a finite set and $\mathcal{D}=\{D_n\}_{n>0}$ be a 
parametrized family of domains. 
$\mathcal{R}^{\mathcal{D}}$ denote the set of all functions
mapping from $\mathcal{D}$ to $\mathcal{R}$. A \emph{property $\proP$}
is simply specified by a family of functions $\proP\subseteq\mathcal{R}^{\mathcal{D}}$.
A \emph{tester} for property $\mathcal{P}$ is a randomized algorithm
which, given the oracle access to an input function $f\in\mathcal{R}^{\mathcal{D}}$
together with a distance parameter $\eps$, distinguishes with high
probability (say, $2/3$) between the case that $f$ satisfies $\proP$
and the case that $f$ is \emph{$\eps$-far} from satisfying $\proP$.
Here, distance between functions $f,g:\mathcal{D}\to\mathcal{R}$,
denoted $\text{dist}(f,g)$, is simply the probability that $\Pr_{x\in\mathcal{D}}[f(x)\ne g(x)]$,
where $x$ is chosen uniformly at random from $\mathcal{D}$, and
$\text{dist}(f,\proP)=\min_{g\in\proP}\{\text{dist}(f,g)\}$. We say
$f$ is \emph{$\eps$-far} from $\proP$ if $\text{dist}(f,\proP)\geq\eps$
and \emph{$\eps$-close} otherwise. The central parameter associated
with a tester is the number of oracle queries it makes to the function
$f$ being tested.

Property testing was first studied by Blum, Luby and Rubinfeld~\cite{BLR93}
and was formally defined by Rubinfeld and Sudan~\cite{RS96}. The
systematic exploration of property testing for combinatorial properties
was initiated by Goldreich, Goldwasser, and Ron~\cite{GGR98}. Subsequently,
a rich collection of properties have been shown 
to be testable~\cite{AS05,AS05b,AFNS06,BCLSSV06,PRS03,AKNS00,AKKLR03,KR04,JPRZ04}.

Perhaps the most fundamental question in property testing is the following:
which properties have {\em local} testing algorithms whose running time depends
only on the distance parameter $\eps$? Are there any attributes that
make a property locally testable? Questions of this type in the
context of graph property testing were first raised in~\cite{GGR98}
and later received a lot of attention. Some very general results have
been obtained~\cite{AFKS00,AS05,AS05b,FN07,AFNS06,BCLSSV06}, leading
to an (almost) complete qualitative understanding of which graph properties
are efficiently testable in the dense graph model (see~\cite{BSS08}
for some recent progress in the sparse graph model). In addition,
for an important class of properties, namely $H$-freeness for fixed
subgraphs $H$, it is known exactly for which $H$, testing $H$-freeness
requires the query complexity to be super-polynomial in $1/\eps$
and for which only a polynomial number of queries suffice: This was
shown by Alon~\cite{Alo02} for one-sided error testers and by Alon
and Shapira~\cite{AS04} for general two-sided error testers. Progress
toward similar understanding has also been made for hypergraph properties~\cite{RS09,AT08,AS05b}.

However, much less is known for algebraic properties. In a systematic study, 
Kaufman and Sudan~\cite{KS08} examined the query complexity
of a broad class of algebraic properties based on the invariance of
these properties under linear transformations. Roughly speaking, they
showed that any locally-characterized linear-invariant and \emph{linear}%
\footnote{A property $\calf$ is linear if for any $f$ and $g$ that are in
$\calf$ necessarily implies that $f+g$ is in $\calf$.%
} properties are testable with query complexities polynomial in $1/\eps$.
Non-linear linear-invariant properties were first shown to be testable
by Green~\cite{Gre05} and were formally studied in~\cite{BCSX09}.
The properties studied in~\cite{Gre05,BCSX09} are ``pattern-freeness''
of Boolean functions, which has been attracting 
considerable attention~\cite{Gre05,BCSX09,Sha09,KSV08,BGS10}, 
as such a study may lead to a complete characterization of testability
for functions, analogous to the setting of graphs. 

\subsection{Motivation for set-theoretic operations}

In this paper we propose a new paradigm to systematically study algebraic property
testing.  First, decompose a natural algebraic property into the union
or intersection (or some other set operation) of a set of  ``atomic properties''. 
Second, try to show that each of these atomic properties is testable. 
Finally, prove that some  ``composite'' property obtained from 
applying some set theoretic operations on the
(testable) atomic properties is also testable. 
A prominent example is the set of low-degree polynomials~\cite{AKKLR03,KR04,JPRZ04}. It is easy
to see that the property of being a degree-$d$ polynomial over GF($2$)
is simply the \emph{intersection} of $2^{2^{d+1}-2}$ atomic properties. 
Indeed, let $\mathbb{P}_{d}$ denote the set of $n$-variate
polynomials of degree at most $d$. Then, by the characterization
of low-degree polynomials (see, e.g., \cite{AKKLR03}), $f\in\mathbb{P}_{d}$
if and only if for every $x_1,\ldots,x_{d+1}\in \F_2^n$, 
\[
\sum_{\emptyset \neq S \subseteq [d+1]} f(\sum_{i\in S} x_i) \equiv 0 \imod{2}.
\]

Now fix an ordering of the non-trivial subsets of $[d+1]=\{1,2,\ldots,d+1\}$. 
Let $\vec{b}$ be a bit-string of length $2^{2^{d+1}-1}$ with an odd number of ones and
$\mathbb{P}_{d,\vec{b}}$ denote the set of functions $f$ such that the string 
$\langle f(\sum_{i\in S} x_i) \rangle_{\emptyset \neq S \subseteq [d+1]}$ is not equal to $\vec{b}$.
By definition, $\mathbb{P}_d$ is the intersection of $2^{2^{d+1}-2}$ ``$\vec{b}$-free'' properties 
$\mathbb{P}_{d,\vec{b}}$'s.%
\footnote{In fact, some of these $2^{2^{d+1}-2}$ properties are identical 
since the set of non-trivial subsets generated by $x_i$ is invariant under permutation of the $x_i$'s.} 

In order to carry out this program of decomposing an algebraic properties into atomic ones, 
one must have a solid understanding of how basic set-theoretic operations affect testability. 
For instance, given two testable properties, is the union, intersection, 
or set-difference also testable? 
Previously, Goldreich, Goldwasser and Ron 
considered such questions in their seminal paper~\cite{GGR98}. 
They observed that the union of two testable properties 
is always testable (cf. Section~\ref{Sec:union}) but
also provided examples showing that in general, testability is not
closed under other set-theoretic operations. 
Thus, current understanding of testability via set-theoretic operations 
seems insufficient to carry out the above mentioned program of attack. 

\subsection{Our results}

In this paper, we show more positive results for these basic set-theoretic
operations and illustrate several applications. We now describe our
contribution in more detail.

\paragraph{Set-theoretic operations}

We provide sufficient conditions that allow local testability to be closed
under intersection and set difference. Given two locally testable properties,
we show that if the two properties (minus their intersection) are
sufficiently far apart, then their intersection is also locally testable.
For set difference, a similar statement can also be made, albeit with
more technicality, requiring that one of the properties must be   ``tolerantly
testable''. 

A more detailed treatment of these set operations appears in Section~\ref{Sec:theory}.
We remark that in the general case, testability is not closed under
most set operations. Thus, putting restrictions on these properties
is not unwarranted. 

Applications of these set-theoretic considerations appear in Sections~\ref{Sec:linearity}
and~\ref{Sec:LB}. Furthermore, Section~\ref{Sec:LB} demonstrates
the simplicity that comes from these set-theoretic arguments. 
There, via set theory, we define a new property from an established one, and show that 
the new property's testability, in terms of both upper and lower bounds, is
inherited from the previous property. 

\paragraph{Disjunction of linear functions}
In addition to set theory, it is also natural to ask whether testability 
is preserved under the closure of some fundamental, unary operations. 
For instance, given a testable property $\proP$, 
under what condition is its additive closure $\oplus \proP$ testable? 
A similar question can also be asked 
for the disjunctive operator $\wedge$, 
which is one of the most basic operations used to combine formulas. 
Given a testable property $\proP$, is its disjunctive closure $\wedge \proP$ testable? 

Trivially, if $\proP$ is linear, then $\oplus \proP=\proP$ and testability is preserved.
Furthermore, if $\proP_1$ and $\proP_2$ are both linear 
and linear-invariant as introduced by Kaufman and Sudan~\cite{KS08}, 
then their sumset $\proP_1+\proP_2$ is testable. 
However, in general, not much can be said 
about how these basic operations affect testability.

Here we focus on disjunction's effect on one specific property, namely the set of linear functions. 
Before we describe our result, we note some previous works in testing where disjunction played a role. 
For the disjunction of monomials,
Parnas et. al.~\cite{PRS03} gave a testing algorithm for $s$-term monotone DNF with
query complexity $\tilde{O}(s^{2}/\eps)$. 
Diakonikolas et. al.~\cite{DLM+07} generalized Parnas et. al.'s result
to general $s$-term DNF with query complexity $\tilde{O}(s^{4}/\eps^{2})$.  

We take a different direction and ask how disjunction affects 
the testability of the set of linear functions.
The property of being a linear Boolean function (see next section for a full discussion), 
first studied by Blum, Luby and Rubinfeld~\cite{BLR93}, 
is testable with query complexity $O(1/\eps)$.
As observed in~\cite{BCSX09}, 
the class of disjunction of linear functions 
is equal to the class of $100$-free functions (see Preliminaries for a definition). 
There they showed that a sufficiently rich class of ``pattern-free'' functions 
is testable, albeit with query complexity a tower of $2$'s whose height is a function of $1/\eps$.
In a different context, the authors in~\cite{GOSSW} showed implicitly%
\footnote{We thank an anonymous reviewer from ICS 2011 for pointing this out.}
that the disjunction of linear functions 
is testable with query complexity polynomial in $1/\eps$, but with two-sided error.

Since both \cite{BCSX09} and \cite{GOSSW} seek to describe rich classes of testable Boolean functions, 
the bounds from both works do not adequately address how disjunction 
affects the query complexity of the underlying property, the set of linear functions. 
In Section~\ref{Sec:100-free}, we give a direct proof, 
showing that the disjunction of linear functions is testable 
with query complexity $O(1/\eps^2)$ and has one-sided error. 
Thus, the blowup from the disjunctive operator is $O(1/\eps)$. 
It will be interesting to see if the blowup is optimal for this problem.

\paragraph{A different proof for linearity testing}
Linearity testing, first proposed by Blum, Luby and Rubinfeld~\cite{BLR93},
is arguably the most fundamental and extensively studied problem in
property testing of Boolean functions. Due to its simplicity and important
applications in PCP constructions, much effort has been devoted to
the study of the testability of linearity~\cite{BLR93,BGLR93,BGS98,BCHKS96,Kiw03}.

For linearity, we indeed are able to carry out the program of decomposing
an algebraic property into atomic pattern-free properties, and thus
obtain a novel new proof that linearity is testable in Section~\ref{Sec:linearity}.
In particular, linearity is easily seen to be equal to the intersection
of two atomic properties, namely triangle-freeness (see Section~\ref{Sec:prelim})
and disjunction of linear functions, which are both testable. 

The query complexity of linearity in our proof is of the tower-type,
drastically worse than the optimal $O(1/\eps)$ bound, where $\eps$
is the distance parameter. 
We note that our effort in obtaining a new proof lies not in improving the parameters, 
but in understanding
the relationships among these atomic, testable properties. 
In fact, we believe that despite the poor upper bound, our new proof is conceptually
simple and gives evidence that set theory may uncover new testable
properties.

\subsection{Techniques}

Our new proof that linearity is testable is built on the testability
results for triangle freeness (see definition in Section~\ref{Sec:prelim})
and the disjunction of linear functions. 
The latter was already shown to be testable in~\cite{BCSX09}.
However, in this work, we give a completely different proof using a BLR-styled approach. 
Our proof is a novel variant of the classical self-correction method. 
Consequently, the query upper bound we obtain (quadratic in $1/\eps$) is significantly
better than the tower-type upper bound shown in~\cite{BCSX09}. 
In fact, to the best of our knowledge, this is the first and only polynomial query 
upper bound for testing pattern-freeness properties. 
All other analysis for testing pattern-freeness properties apply some type of ``regularity lemma'', 
thus making tower-type query upper bounds unavoidable. 

We believe that both the self-correction technique and the investigation of set-operations 
may be useful in the study of testing pattern-freeness. 
From the works developed in~\cite{Sha09,KSV08}, 
we know that for every $d$, the property $\mathbb{P}_{d,\vec{1}}$ is testable.%
\footnote{Actually, stronger theorems were proved in~\cite{Sha09,KSV08}, 
but to state their works in full, definitions not needed in this work will have to be introduced.} 
However, for an arbitrary $\vec{b}$, 
the testability of $\mathbb{P}_{d,\vec{b}}$ remains open. 
And in general very little can be said about the testability 
of an arbitrary intersection of these properties.  
Since $\mathbb{P}_{d}$ 
is known to be testable using self-correction~\cite{AKKLR03}, 
we believe that self-correction, 
applied in conjunction with set-theory, 
may be useful for understanding these pattern-free properties. 


\section{\texorpdfstring{Preliminaries}{2. Preliminaries}}

\label{Sec:prelim} 
We now describe some basic notation and definitions that we use throughout the paper.
We let $\N=\{0,1,\ldots\}$ denotes the set of natural numbers 
and $[n]$ the set $\{1,\ldots,n\}$. 
We view elements in $\ncube$ as $n$-bit binary strings, that is elements of $\{0,1\}^{n}$.
For $x \in \ncube$, we write $x_{i}\in \{0,1\}$ for the $i^{\text{th}}$ bit of $x$.
If $x$ and $y$ are two $n$-bit strings, then $x+y$ denotes bitwise
addition (i.e., XOR) of $x$ and $y$,
and $x \cdot y=\sum_{i=1}^{n}x_{i}y_{i}\pmod{2}$ denotes the inner 
product between $x$ and $y$.
We write $(x,y)$ to denote
the concatenation of two bit strings $x$ and $y$. For convenience,
sometimes we view a $n$-bit binary string as a subset of $[n]$,
that is, for every $x\in\ncube$ there is a corresponding subset $S_{x}\subseteq[n]$
such that $x_{i}=1$ iff $i\in S_{x}$ for every $1\leq i\leq n$.
We write $|x|$ to indicate the Hamming weight of $x$, i.e., the
number of coordinates $i$ such that $x_{i}=1$. Equivalently, this
is also the cardinality of subset $S_{x}$.
By abuse of notation, we use parentheses to denote
\emph{multisets}; for instance, 
we write $(a, a, b, b, b)$ for the multiset which consists of two $a$'s
and three $b$'s. 

Let $f:\ncube\to\cube$ be a Boolean function. The \emph{support}
of $f$ is $\supp(f)=\{x\in\ncube:f(x)=1\}$. Recall that for two
functions $f$ and $g$ defined over the same domain, the (fractional) \emph{distance}
between these two functions is $\dist(f,g)\eqdef\Pr_{x\in\mathcal{D}}[f(x)\ne g(x)]$.
Let $\proPo$ and $\proPt$ be two properties defined over the same
domain $\mathcal{D}$, then the \emph{distance} between these two
properties, $\text{dist}(\proPo,\proPt)$, is simply defined to be
$\min_{f\in\proPo,g\in\proPt}\{\text{dist}(f,g)\}$.

A Boolean function $f:\ncube\to\cube$ is \emph{linear} if
for all $x$ and $y$ in $\ncube$, $f(x)+f(y)=f(x+y)$. We denote
the set of linear function by $\lin$. Throughout this paper, we will
be working with the pattern generated by the triple $(x,y,x+y)$.
To this end, we say that a Boolean function $f:\ncube\to\cube$ is
\emph{$(1,0,0)$-free} if for all $x$ and $y$ in $\ncube$, $(f(x),f(y),f(x+y))\neq (1,0,0)$,
where here and after we view $(f(x),f(y),f(x+y))$ as well as $(1,0,0)$
as multisets%
\footnote{That is, for example, 
we do not distinguish the case $\langle f(x),f(y),f(x+y)\rangle=\langle1,0,0\rangle$
from $\langle f(x),f(y),f(x+y)\rangle=\langle0,1,0\rangle$.%
}. 
We denote the set of $(1,0,0)$-free functions by $\ozz$.
Similarly, a\emph{ $(1,1,0)$-free} Boolean function is defined analogously.
Lastly, we say that a Boolean function $f:\ncube\to\cube$ is \emph{triangle-free}
if for all $x$ and $y$ in $\ncube$, $(f(x),f(y),f(x+y))\neq (1,1,1)$.
We denote the set of triangle-free functions by $\ooo$. Note that
$\ooo$ is \emph{monotone}: if $f\in\ooo$ and we modify $f$ by setting
some of the points in $\ncube$ from $1$ to $0$, then the new function
is clearly also triangle-free. 
We encapsulate this observation into the following statement:
\begin{obs}\label{obs:monotone}
Let $f$ and $g$ be two Boolean functions such that $\supp(f)\subseteq\supp(g)$. 
Then \[\dist(f,\ooo)\leq \dist(g,\ooo).\]
\end{obs}

For concreteness, we provide a formal definition of a tester.

\begin{defn}[Testability] 
Let $\mathcal{R}$ be a finite set and 
$\mathcal{D}=\{D_n\}_{n>0}$ be a parametrized family of domains. 
Let $\proP\subseteq\mathcal{R}^{\mathcal{D}}$ be a property. 
We say a (randomized) algorithm $T$ is a \emph{tester} for $\proP$
with \emph{query complexity} $q(\eps,n)$ 
if for any distance parameter $\eps>0$, input size $n$ 
and function $f:D_n \rightarrow R$,
$T$ satisfies the following:

\begin{itemize}
\item $T$ queries $f$ at most $q(\eps,n)$ times;
\item (completeness) if $f\in\proP$, then $\Pr[\text{\ensuremath{T} accepts}]=1$;
\item (soundness) if $\dist(f,\proP)\geq\eps$, 
then $\Pr[\text{\ensuremath{T} accepts}]\leq \frac{1}{3}$,
where the probabilities are taken over the internal randomness used by $T$.
\end{itemize}
\end{defn}

We say that a property is \emph{locally testable} 
if it has a tester whose query complexity is a function depending only on $\eps$, 
independent of $n$. 
In this work, we actually use the word testability to 
describe the stronger notion of \emph{local testability}. 
For our main results, we will work with the model case when 
$\mathcal{D}_n=\ncube$ and $\mathcal{R}=\{0,1\}$. 

\section{\texorpdfstring{Basic theory of set operations}{3. Basic theory of set operations}}
\label{Sec:theory}


In this section, we present some basic testability results based on
set-theoretic operations such as union, intersection, complementation,
and set-difference. The proofs here are fairly standard and are 
thus deferred to the Appendix.

\subsection{Union}
\label{Sec:union} 
It is well known that the union of two testable properties remains testable. 
This folklore result first appeared in~\cite{GGR98};
for completeness, a proof is included in Appendix~\ref{Sec:union proof}.

\begin{prop}[Folklore]\label{prop:OR} 
Let $\proPo,\proPt\subseteq\mathcal{R}^{\mathcal{D}}$ 
be two properties defined over the same domain $D=\{D_n\}_{n>0}$. 
For $i=1,2$, suppose
$\mathcal{P}_{i}$ is testable with query complexity $q_{i}(\eps)$. 
Then the union $\proPo\cup\proPt$ is testable with query complexity $O(q_{1}(\eps)+q_{2}(\eps))$.
\end{prop}


\subsection{Intersection}

\label{Sec:intersection} 
The case of set intersection is more complicated than union. 
Goldreich et al. showed in~\cite{GGR98} (see Proposition 4.2.2) 
that there exist testable properties whose intersection is not testable. 
Thus, in general, testability does not follow from the intersection operation. 
However, testability may still follow in restricted cases. 
In particular, we show that if two testable properties $\proPo$
and $\proPt$ minus their intersection are sufficiently far from each
other, then their intersection remains testable as well. 
A proof is included in Appendix~\ref{sec:intersection proof}.

\begin{prop}\label{prop:AND} 
Let $\proPo,\proPt\subseteq\mathcal{R}^{\mathcal{D}}$
be two properties defined over the same domain $\mathcal{D}=\{D_n\}_{n>0}$. 
Suppose $\text{dist}(\proPo\setminus\proPt,\proPt\setminus\proPo)\geq\eps_{0}$
for some absolute constant $\eps_{0}$, and for $i=1,2$, $\mathcal{P}_{i}$
is testable with query complexity $q_{i}(\eps)$.
Then the intersection $\proPo\cap\proPt$ 
is testable with query complexity $O(q_{1}(\eps)+q_{2}(\eps))$, 
\end{prop} 

\subsection{Complementation}

\label{Sec:complement} Here we examine the effect complementation
has on the testability of a property. 
As it turns out, all three outcomes -- both $\mathcal{P}$ and $\bar{\proP}$ are testable, only one of
$\mathcal{P}$ and $\bar{\proP}$ is testable, and neither $\mathcal{P}$
nor $\bar{\proP}$ is testable -- are possible! 

The first outcome is the easiest to observe. Note that the property
$\mathcal{D^{R}}$ and the empty property are complements of each
other, and both are trivially testable. The second outcome is observed
in~Proposition 4.2.3 in~\cite{GGR98}. To our knowledge, the third
outcome has not been considered before. In fact, previous constructions
of non-testable properties, e.g.~\cite{GGR98,BHR05}, are \emph{sparse}.
Hence, the complements of these non-testable properties are trivially
testable (by the tester that accepts all input functions). One may
wonder if in general the complement of a non-testable property must
also be testable. We disprove this in the following proposition. 

\begin{prop}\label{prop:complement}
There exists some property $\proP\subseteq\mathcal{R}^{\mathcal{D}}$ where 
$\mathcal{R}=\{0,1\}$ and $\mathcal{D}=\{\ncube\}_{n>0}$, 
such that neither $\proP$ nor $\overline{\proP}$
is testable for any $\eps<1/8$.
\end{prop} 

By utilizing coding theory, 
we can bypass the sparsity condition to prove Proposition~\ref{prop:complement}.
Essentially, property $\proP$ consists of neighborhoods 
around functions that have degree $n/2-1$ as polynomials over $\F_{2^n}$. 
Its complement contains functions that are polynomials of degree $n/2$. 
Since $d$ evaluations are needed to specify 
a polynomial of degree $d$, 
any tester for $\proP$ or $\overline{\proP}$ needs (roughly) at least $n/2$ queries. 
Using a standard argument involving code concatenation, 
one can construct $\proP$ and $\overline{\proP}$ to be 
\emph{binary} properties that require testers of
query complexity $\Omega(2^{n/2})$. 
A formal proof can be found in Appendix~\ref{Sec:complement proof}.


\subsection{Difference}

\label{Sec:difference} 
Let $\proPo$ and $\proPt$ be two properties
and let $\proP=\proPo\setminus\proPt$ denote the set difference of
the two properties. In this section, we confine our attention to the
simple case that $\proPt\subset\proPo$. Since complementation is
a special case of set-difference, from Section~\ref{Sec:complement},
we know that in general we can infer nothing about the testability
of $\proP$ from the fact that both $\proPo$ and $\proPt$ are testable.
However, under certain restrictions, we still can show that $\proP$
is testable.

First we observe a simple case in which $\proPo\setminus\proPt$ is
testable. This simple observation, which is obvious and whose proof
we omit, is utilized in the proof of Theorem~\ref{thm:LB} in Section~\ref{Sec:LB}.

\begin{obs}\label{obs:set-difference} 
Let $\proP_2 \subset \proP_1$ be two testable properties 
defined over the same domain $\mathcal{D}=\{D_n\}_{n>0}$.
If for every $f\in\proPt$, 
there is some $g\in\proP_1\setminus \proP_2$ such that $\dist(f,g)=o(1)$,
then $\proP_1\setminus \proP_2$ is testable by the same tester which tests property
$\proPo$. 
\end{obs} 

Our second observation on set difference relies on the notion of \emph{tolerant
testing}, introduced by Parnas, Ron, and Rubinfeld~\cite{PRR06}
to investigate testers that are guaranteed to accept (with high confidence)
not only inputs that satisfy the property, but also inputs that are
sufficiently close to satisfying it. 

\begin{defn}[Tolerant Tester~\cite{PRR06}]
Let $0<\eps_{1}<\eps_{2}<1$ denote two distance parameters and
$\proP\subseteq\mathcal{R}^{\mathcal{D}}$ be a property defined 
over the domain $\mathcal{D}=\{D_n\}_{n>0}$.
We say that property $\proP$ is $(\eps_1,\eps_2)$-\emph{tolerantly testable} 
with query complexity $q(\eps_1,\eps_2)$ 
if there is a tester $T$ that makes at most $q(\eps_1,\eps_2)$ queries,
if for all $f$ with $\dist(f,\proP)\leq\eps_{1}$, $T$ rejects $f$
with probability at most $1/3$, and for all $f$ with $\dist(f,\proP)\geq\eps_{2}$,
$T$ accepts $f$ with probability at most $1/3$. 
\end{defn}

We record in the following proposition that if $\mathcal{P}$ and
$\mathcal{P}_{2}$ are sufficiently far apart and $\mathcal{P}_{2}$
is tolerantly testable, then $\mathcal{P}$ is also testable. We include
a proof in Appendix~\ref{Sec:set-difference}.

\begin{prop}\label{thm:set-difference} 
Let $\eps_{1}<\eps_{2}<\eps_{0}$ be three absolute constants.
Let $\proP_{2}\subset\proP_{1}\subseteq\mathcal{R}^{\mathcal{D}}$ 
be two properties defined over the same domain $\mathcal{D}=\{D_n\}_{n>0}$.
If for every $\eps>0$, $P_1$ is testable with query complexity $q_1(\eps)$, 
$P_2$ is $(\eps_1,\eps_2)$-tolerantly testable with query complexity $q_{2}(\eps_1,\eps_2)$, 
and $\dist(\proP_1\setminus \proP_2,\proP_{2})\geq\eps_{0}$, 
then $\proP_1\setminus \proP_2$ is testable 
with query complexity $O(q_1(\eps)+q_2(\eps_1,\eps_2))$ (and completeness $2/3)$.
\end{prop} 

We note that since $\proP_2$ is tolerantly testable, 
it does not have completeness $1$. 
Thus, the set difference $\proP_1 \setminus \proP_2$ 
is not guaranteed to have one-sided error, either.


\section{\texorpdfstring{Main results}{4. Main results}}

\label{Sec:applications} In this section we show two applications
of the results developed in Section~\ref{Sec:theory}. We stress
that set theoretic arguments may be used to show both upper bound
results (some properties are testable with only a few number of queries)
and lower bound results (some properties can not be tested by any
tester with less than certain number of queries).

\subsection{Testing disjunction of linear functions}
\label{Sec:100-free} 
In this section, we employ a BLR-style analysis to show that the class 
of disjunction of linear functions is testable with query complexity $O(1/\eps^2)$.
We first recall from~\cite{BCSX09} that a function is a disjunction of linear functions 
iff it is $(1,0,0)$-free. (Recall that $\ozz$ is the set of Boolean functions
that are free of $(1,0,0)$-patterns for any $x$, $y$ and $x+y$ in
$\ncube$.)

\begin{prop}[\cite{BCSX09}]\label{prop:char} 
A function $f:\ncube\rightarrow\{0,1\}$ is (1,0,0)-free if and only if $f$ is
the disjunction (OR) of linear functions (or the all $1$ function).
\end{prop} 
\begin{proof} 
The reverse direction is obvious. 
For the forward direction, let $S=\{x\in\ncube:f(x)=0\}$. 
If $S$ is empty, then $f$ is the all $1$ function. 
Otherwise let $x$ and $y$ be any two elements in $S$ (not necessarily distinct). 
Then if $f$ is $(1,0,0)$-free, it must be the case that $x+y$ is also in
$S$. Thus $S$ is a linear subspace of $\ncube$. 
Suppose the dimension of $S$ is $k$ with $k\geq1$. 
Then there are $k$ linearly independent vectors $a_{1},\ldots,a_{k}\in\ncube$ such that $z\in S$ 
iff $\{z\cdot a_{1}=0\}\bigwedge\cdots\bigwedge\{z\cdot a_{k}=0\}$.
Therefore, by De Morgan's law, $f(z)=1$ 
iff $z\in\bar{S}$ iff $\{z\cdot a_{1}=1\}\bigvee\cdots\bigvee\{z\cdot a_{k}=1\}$,
which is equivalent to the claim. \end{proof}

$\ozz$ was shown to be testable with a tower-type query
upper bound in~\cite{BCSX09}. 
We now give a direct proof that $\ozz$ is testable with a quadratic upper bound. 
In fact, by symmetry the testability of $\ooz$ is the same as the testability of $\ozz$. 

\begin{theorem}\label{thm:100} 
For every distance parameter $\eps>0$, the property $\ozz$ is testable 
with query complexity $O(1/\eps^2)$.
\end{theorem} 
\begin{proof} 
Suppose we have oracle access to some Boolean function $f:\ncube\rightarrow\cube$. 
A natural $3$-query test $T$ for $\ozz$ proceeds as follows. 
$T$ picks $x$ and $y$ independently and uniformly at random from $\ncube$, 
and accepts iff $(f(x),f(y),f(x+y))\neq (1,0,0)$. 

Let $R\eqdef\Pr_{x,y}[(f(x),f(y),f(x+y))\neq (1,0,0)]$ be the rejection probability of $T$.
If $f\in \ozz$, then $R=0$, i.e., $T$ has completeness $1$. 
For soundness, in a series of steps, we shall show that  
for every $\eps>0$, if $R<\eps^2/128$, then there exists 
a Boolean function $g$ such that {\bf (1)} $g$ is well-defined, {\bf (2)} $\dist(f,g)<\eps$, and 
{\bf (3)} $g$ is in $\ozz$.

Let $\mu_0$ denote $\Pr_{x}[f(x)=0]$. 
Suppose $\mu_0 < 63\eps/64$. Then $\dist(f,\vec{1})<63\eps/64$, 
where $\vec{1}$ is the all-ones function. 
Then trivially, taking $g=\vec{1}$ completes the proof.  
Thus, henceforth we assume that $\mu_0 \geq 63\eps/64$.
\vskip10pt
For a fixed $x\in\ncube$, let $p_{00}^{x}$ denote $\Pr_{y}[(f(y),f(x+y))=(0,0)]$, and 
and $p_{10}^{x}$ is defined similarly. We define $g:\ncube\rightarrow\cube$ as follows: 
\[
g(x)=\begin{cases}
0, & \text{ if $p_{00}^{x}\geq\eps/4$;}\\
1, & \text{ if $p_{10}^{x}\geq\eps/4$;}\\
f(x), & \text{ otherwise.}\end{cases}\]

\paragraph{Proof of (1).} $g$ is well-defined. 

Suppose not, then there exists some $x\in\ncube$ such that $p_{00}^{x},p_{10}^{x}\geq\eps/4$.
Pick $y$ and $z$ independently and uniformly at random from $\ncube$.
Let $E$ be the event that 
\[
\text{at least one of } (f(y),f(z),f(y+z)) \text{ and } 
(f(x+y),f(x+z),f(y+z)) \text{ is } (1,0,0).
\]

By assumption, with probability at least $\eps^2/16$, $f(y)=1,f(x+y)=0$ and $f(z),f(x+z)=0$, 
which will imply that -- regardless of the value of $f(y+z)$ -- event $E$ must occur.
Thus, $\eps^2/16\leq\Pr[E]$. 
On the other hand, by the union bound, $\Pr[E]\leq2R<\eps^2/64$, a contradiction.
\qed

\paragraph{Proof of (2).} $\dist(f,g)< \frac{\eps}{32}$.

Suppose $x$ is such that $f(x)\neq g(x)$.
By construction, $\Pr_y[f(x),f(y),f(x+y)]\geq \eps/4$.
This implies that the rejection probability $R$ is at least $\dist(f,g)\cdot \eps/4$. 
Since $R<\eps^2/128$, $\dist(f,g)<\eps/32$.
\qed

\vskip13pt
Before proving {\bf(3)}, we first note that for every $x\in \ncube$,
\label{eq:g low violation}
\[\Pr_{y}[(g(x),g(y),g(x+y))=(1,0,0)]<\frac{5\eps}{16}.
\]
To see this, note that by construction of $g$, 
for every $x\in\ncube$, $\Pr_{y}[(g(x),f(y),f(x+y))=(1,0,0)]<\eps/4$.
Since $\dist(f,g)<\eps/32$, by the union bound, we can deduce that the probability that 
$g$ has a $(1,0,0)$-pattern at $(x,y,x+y)$ is less than $\eps/4+2\cdot \eps/32$.

\paragraph{Proof of (3).} $g$ is in $\ozz$.

Suppose not, that there exist $x,y\in \ncube$ such that $g(x)=1$, $g(y),g(x+y)=0$.
Pick $z$ uniformly at random from $\ncube$.
Let $E$ denote the event that 
\begin{align*}
&\text{at least one of } (g(x),g(z),g(x+z)), (g(y),g(z), g(y+z)),\\
&\quad \text{ and } (g(x+y),g(x+z),g(y+z))  \text{ is } (1,0,0).
\end{align*}
A case by case analysis reveals that if $g(z)=0$, then event $E$ must occur. 
Note that the probability that $g(z)=0$ is at least $63\eps/64-\eps/32=61\eps/64$,
since $f(z)=0$ occurs with probability at least $63\eps/64$ and $\dist(f,g)<\eps/32$.
On the other hand, by union bound, we have $\Pr[g(z)=0]\leq \Pr[E]\leq 3\cdot 5\eps/16$, 
implying that 
$61\eps/64 \leq 15\eps/16$, an absurdity.

Therefore, we have shown that on any input function that is $\eps$-far from $\ozz$,
the rejection probability of $T$ is always at least $\eps^{2}/128$. 
By repeating the basic test $T$ independently $O(1/\eps^{2})$ times, 
we can boost the rejection probability of $T$ to $2/3$, and thus completing the proof. 
\end{proof}

\subsection{A new proof that linearity is testable}
\label{Sec:linearity} %
 
As an application of our results in Section~\ref{Sec:intersection},
we give a new proof that linear functions are testable based on a
set-theoretic argument. 
To this end, note that the set of linear functions equals to the intersection of $(1,1,1)$-free
functions and $(1,0,0)$-free functions, i.e., 
\[\lin=\ooo\cap\ozz.\]
From the previous section, we know that $\ozz$ is testable. The following
theorem due to Green~\cite{Gre05} asserts that $\ooo$ is also testable.

\begin{theorem}[\cite{Gre05}]\label{thm:Green} 
The property $\ooo$ is testable with query complexity $W(\poly(1/\eps))$,
where for every $t>0$, $W(t)$ denotes the tower of $2$'s of height
$\lceil t\rceil$.\end{theorem}
By Proposition~\ref{prop:AND}, to show that linearity is testable,
it suffices to show that the two properties $\ooo$ and $\ozz$ are
essentially far apart. 
To this end, let us define a new property $\LBP$, 
where NLTF stands for \emph{non-linear triangle-freeness}:
\[\LBP\eqdef\ooo\setminus\lin.\]
\begin{lemma}\label{lemma:distance} 
We have that
$\ozz\setminus\lin$ is $\frac{1}{4}$-far from $\LBP$. 
\end{lemma} 
We first establish a weaker version of Lemma~\ref{lemma:distance}.
\begin{prop}\label{prop:distance-2-disj}
Suppose $f$ is a disjunction of exactly two non-trivial linear functions. 
Then $\dist(f,\ooo)$ is at least $\frac{1}{4}.$ 
\end{prop}
\begin{proof} 
Set $N=2^{n}$. Write $f(x)=(\alpha\cdot x)\bigvee(\beta\cdot x)$, where 
$\alpha\neq\beta\in\ncube$ denote two $n$-bit vectors not equal to $0^{n}$.
We say that a tuple $(x,y,x+y)$ where $x,y\in\ncube$ is a triangle in $f$ if $f(x),f(y),f(x+y)=1$.
We shall show that
{\bf (1)} $f$ has $N^2/16$ triangles and 
{\bf (2)} for every $x$, the number of $y's$ such that $(x,y,x+y)$ is a triangle in $f$ is $N/4$.
Together, {\bf (1)} and {\bf (2)} will imply that $\dist(f,\ooo)$ is at least $1/4$, 
since changing the value of $f$ at one point removes at most $N/4$ triangles. 

To prove these two assertions, let $A=\{x\in\ncube:\alpha\cdot x=1\}$ 
and $B=\{x\in\ncube:\beta\cdot x=1\}$.
Since $\supp(f)=A\cup B$, for every triangle $(x,y,x+y)$ in $f$, 
each of the three points $x,y,x+y$ must fall in one of the following three disjoint sets:
\[A\setminus B, (A\cap B), B\setminus A.\]

Furthermore, each of the three points must fall into distinct sets. 
To see this, suppose that $x,y \in A\setminus B$. 
Then by definition, $\alpha(x+y)=\alpha(x)+\alpha(y)=0$ and $\beta(x+y)=0$, 
implying that $f(x+y)=0$, a contradiction. 
So $A\setminus B$ cannot contain two points of a triangle, and by symmetry, neither can $B\setminus A$.
The same calculation also reveals that $A\cap B$ cannot contain two points of a triangle.

Thus, a triangle $(x,y,x+y)$ in $f$ must be such that 
$x\in A\setminus B$, $y \in A\cap B$, and $x+y \in B\setminus A$.
In addition, it is easy to check that 
given two points $p_1$, $p_2$ from two distinct sets (say $A\setminus B$ and $A\cap B$), 
their sum $p_1+p_2$ must be in the third set ($B\setminus A$). 
Since these three sets $A\setminus B, (A\cap B), B\setminus A$ all have size $N/4$, 
this implies that 
the number of triangles in $f$ is $N^2/16$, proving {\bf (1)}.

{\bf (2)} also follows easily given the above observations. 
Suppose $x \in A\setminus B$. For every $y\in A\cap B$, $(x,y,x+y)$ forms a triangle.
Since any triangle that has $x$ as a point 
must also contain a point in 
$A \cap B$ (with the third point uniquely determined by the first two), 
the number of triangles in $f$ containing $x$ is $N/4$. 
The case when $x\in B\setminus A$ or $x\in A\cap B$ is similar. This completes the proof. 
\end{proof}
Now we prove Lemma~\ref{lemma:distance}. 
\begin{proof}[Proof of Lemma~\ref{lemma:distance}]
Let $f\in \ozz\setminus\lin$ and write $f=f_1 \vee f_2$, 
where $f_1$ is a disjunction of exactly two linear functions.  
By Proposition~\ref{prop:distance-2-disj}, it follows that $\dist(f_1,\ooo)$ is at least $1/4$. 
Since $\ooo$ is monotone and $\supp(f_1)\subseteq \supp(f)$, 
by Observation~\ref{obs:monotone}, we know 
that $\dist(f,\ooo)\geq 1/4$. 
Since $\LBP \subset \ooo$, $\dist(f,\LBP)\geq\dist(f,\ooo)$, completing the proof. 
\end{proof}

By Theorem~\ref{thm:Green} and Theorem~\ref{thm:100}, both $\ooo$
and $\ozz$ are testable. Now by combining Proposition~\ref{prop:AND}
and Lemma~\ref{lemma:distance}, we obtain the following: 
\begin{theorem}\label{thm:linearity}
$\lin$ is testable. 
\end{theorem}

We remark that the query complexity for testing linearity in Theorem~\ref{thm:linearity}
is of the tower type (of the form $W(\poly(1/\eps))$ because of Theorem~\ref{thm:Green}.
This is much worse than the optimal linear query upper bound obtained
in~\cite{BLR93,BCHKS96}.

\subsection{A lower bound for testing non-linear triangle-freeness}
\label{Sec:LB} 
We first show that $\lin$ is a ``thin strip'' around $\LBP$. 

\begin{prop}\label{prop:thin strip}
For any Boolean function $f$, $\dist(f,\ooo)\geq\dist(f,\LBP)-2^{-n}$.
\end{prop}
\begin{proof}
The statement is trivially true if $\dist(f,\ooo)=\dist(f,\LBP)$.
Since $\LBP$ is a proper subset of $\ooo$, 
we can assume that $\dist(f,\LBP)$ is strictly larger than 
$\dist(f,\ooo)$, implying that the function in $\ooo$ that has minimum distance to $f$ is actually 
in $\lin$. Call this function $g$. 
Then it is easy to see that there exists some function $h$ in $\LBP$ 
such that $\dist(g,h)=2^{-n}$. 
To this end, note that if $g$ is the all-zero function, 
we can define $h$ such that $h(x)=1$ for some $x\neq 0^n$ and $0$ everywhere else. 
By construction $h$ is non-linear but triangle-free. 
If $g$ is a non-trivial linear function, then we can pick any $x\in\supp(g)$ and define 
$h(x)=0$ and $h(y)=g(y)$ for all $y\neq x$. 
By construction $h$ is non-linear, and since $\ooo$ is monotone, $h$ remains triangle-free.

Thus, by Triangle inequality, we know that $\dist(f,\ooo)=\dist(f,g)$ is at least 
$\dist(f,h)-2^{-n}$. 
This implies that $\dist(f,\ooo)\geq\dist(f,\LBP)-2^{-n}$.
\end{proof}
Since any linear function is $2^{-n}$-close to a function
in $\LBP$, intuitively we expect $\LBP$, which is obtained by deleting
the strip $\lin$ from $\ooo$, to inherit the testability features
of $\ooo$. Indeed, we record this next by using the set-theoretic
machinery set up in Section~\ref{Sec:theory}.

\begin{theorem}\label{thm:LB} 
$\LBP$ is testable, but any non-adaptive
\footnote{A tester is non-adaptive if all its query points 
can be determined before the execution of the algorithm, 
i.e., the locations where a tester queries do not depend on the answers to previous queries.}
tester (with one-sided error) for $\LBP$ requires $\omega(1/\eps)$
queries. 
\end{theorem}
\begin{proof} 
We first observe that $\LBP$ is testable with one-sided error. 
By Proposition~\ref{prop:thin strip} and Observation~\ref{obs:set-difference},
the testing algorithm for $\LBP$ is simply the same as the tester
for $\ooo$~\cite{Gre05}. 

Next we show that the lower bound for the query complexity of $\LBP$
is the same as $\ooo$. 
As shown in~\cite{BX10}, any one-sided, non-adaptive tester for 
$\ooo$ requires $\omega(1/\eps)$ queries.%
\footnote{The specific lower bound shown in~\cite{BX10} is 
$\Omega((\frac{1}{\eps})^{1.704\cdots})$ but can be 
improved to be $\Omega((\frac{1}{\eps})^{2.423\cdots})$ 
as observed independently by Eli Ben-Sasson and the third author of the present paper.}
Suppose $\LBP$ is testable with one-sided error and has query complexity
$O(1/\eps)$.
Since $\lin$ is testable with
query complexity $O(1/\eps)$~\cite{BLR93}, by Proposition~\ref{prop:OR} $\ooo=\lin \cup \LBP$
is testable with one-sided error and has query complexity $O(1/\eps)$,
a contradiction. \end{proof}

\section{\texorpdfstring{Concluding remarks}{5. Concluding remarks}}
\label{Sec:conclusion} 
We have initiated a general study of the closure
of testability under various set operations.  Our results show that
such a study can lead to both upper and lower bound results in property
testing. We believe our answers are far from complete, and further
investigation may lead to more interesting results. For example, the
symmetric difference between two properties $\proPo$ and $\proPt$
is defined to be $\proPo\bigtriangleup\proPt\eqdef(\proPo\setminus\proPt)\cup(\proPt\setminus\proPo)$.
Under what conditions is the property $\proPo\bigtriangleup\proPt$
testable if both $\proPo$ and $\proPt$ are testable? Another natural
generalization of our approach is to examine properties resulting
from a finitely many application of some set-theoretic operations. 

Our proof that the class of disjunction of linear functions is testable employs a BLR-style
self-correction approach. 
We believe that this technique may be useful in analyzing other non-monotone, pattern-free properties. 
In particular, it will be interesting to carry out our approach of decomposing
an algebraic property into atomic ones for higher degree polynomials. 
This will, in addition to giving a set-theoretic proof for testing low-degree polynomials, 
sheds light on how pattern-free properties relate to one another.

Finally, our quadratic query complexity upper bound 
for the disjunction of linear functions opens up a number of directions.
In our work, the blowup in query complexity from the disjunction is $O(1/\eps)$. 
One may vary the underlying properties and the operators to measure the blowup in query complexity. 
Of particular interest may be understanding 
how the disjunction affects the testability of low-degree polynomials.

\section*{Acknowledgments}
We thank the anonymous referees for numerous suggestions and the reference to~\cite{GOSSW}.

\bibliographystyle{plain}
\bibliography{papers}

\appendix

\section{Proof of Proposition~\ref{prop:OR}}
\label{Sec:union proof} 
Let $T_{1}$ be the tester for $\proPo$ with query complexity 
$q_{1}(\eps,n)$ and $T_{2}$ be the tester for $\proPt$ with query complexity 
$q_{2}(\eps,n)$. 
We may assume that both $T_1$ and $T_2$ have soundness $1/6$ 
with a constant blowup in their query complexity. 
Define $T$ to be the tester which, on input function $f$, first simulates $T_{1}$ and then $T_{2}$. 
If at least one of the two testers $T_{1}$ and $T_{2}$ accepts $f$, $T$ accepts $f$. 
Otherwise, $T$ rejects.

Clearly the query complexity of $T$ is $O(q_{1}+q_{2})$. 
For completeness, note that if $f$ is in $\proP$, then by definition $f$ is in at least
one of $\proPo$ and $\proPt$. Thus, $T$ accepts $f$ with probability $1$. 
Now suppose $\text{dist}(f,\proP)\geq\eps$. 
Then we have both $\text{dist}(f,\proPo)\geq\eps$
and $\text{dist}(f,\proPt)\geq\eps$. 
By the union bound, the probability that at least one of $T_{1}$ and $T_{2}$ accepts $f$ is at most
$1/6+1/6=1/3$. \qed

\section{Proof of Proposition~\ref{prop:AND}} \label{sec:intersection proof}
Let $T_{1}$ be the tester for $\proPo$ with query complexity $q_{1}(\eps)$, 
and $T_{2}$ be the tester for $\proPt$ with query complexities $q_{2}(\eps)$. 
First we convert $T_{1}$ into another tester $T'_{1}$
for $\proPo$ such that, on input distance parameter $\eps$, $T'_{1}$
makes $Q'_{1}(\eps)$ queries, where
\[
Q_1'(x)=\begin{cases}
q_1(x), & \text{ if $x<\frac{\eps_0}{2}$;}\\
\max\{q_1(x),q_i(\frac{\eps_0}{2})\}, & \text{ otherwise.}
\end{cases}\]
In other words, $T'_1$ can be obtained from $T_1$ by making more queries 
when $x$ is larger than $\eps_0/2$. 
Similarly, we can construct $T'_2$ from $T_2$ in the same manner. 
Since $\eps_0$ is a constant, we have $Q_1'(\eps)=O(q_1(\eps))$ and $Q_2'(\eps)=O(q_2(\eps))$.

Define $T$ to be the tester that on input function $f$, first
simulates $T'_{1}$ and then $T'_{2}$. 
If both testers $T'_{1}$ and $T'_{2}$ accept, then $T$ accepts $f$. 
Otherwise, it rejects. 
The query complexity of $T$ is $Q'_1(\eps)+Q'_2(\eps)$, which is $O(q_{1}(\eps)+q_{2}(\eps))$.

For the completeness, if $f\in\proP$, then both $f\in\proPo$
and $f\in\proPt$ hold. Therefore, $T$ accepts with probability at least $1$. 
For the soundness, suppose $\text{dist}(f,\proP)\geq\eps$. We distinguish
between two cases.

\paragraph{Case 1.} $\eps\leq\frac{\eps_{0}}{2}$. 

It suffices to show that $f$ is $\eps$-far from at least one of $\proP_1$ or $\proP_2$. 
This fact then implies that $T$, in simulating $T'$ and $T_2'$, accepts $f$ with probability at most 
$1/3$. 

To show the $f$ is far from at least one of the two properties, suppose not, 
that we have both $\text{dist}(f,\proPo)<\eps$ and $\text{dist}(f,\proPt)<\eps$. 
That is, there exist $g_{1}\in\proPo$ and $g_{2}\in\proPt$ such that 
$\text{dist}(f,g_{1})<\eps$ and $\text{dist}(f,g_{2})<\eps$.

Since $\text{dist}(f,\proP)\geq\eps$, $g_{1},g_{2}\notin\proP$ and
therefore $g_{1}\in\proPo\setminus\proP$ and $g_{2}\in\proPt\setminus\proP$.
By triangle inequality, $\text{dist}(g_{1},g_{2})<2\eps\leq\eps_{0}$,
and consequently $\text{dist}(\proPo\setminus\proPt,\proPt\setminus\proPo)<\eps_{0}$,
contradicting our assumption. 

\paragraph{Case 2.} $\eps>\frac{\eps_{0}}{2}$. 

There are three sub-cases depending on where $f$ is located. 
We analyze each of them separately below. Note that in each of the sub-cases, 
$f$ is at least $\eps_0/2$-far from one of $\proP_1$ and $\proPt$.
\begin{enumerate}
\item $f\in\proPo\setminus\proP$. 
Then by our assumption on the distance
between $\proPo\setminus\proPt$ and $\proPt\setminus\proPo$, 
$\text{dist}(f,\proPt\setminus\proP)\geq\eps_{0}$.
It follows that 
\begin{align*}
\text{dist}(f,\proPt)
&= \min\{\text{dist}(f,\proP),\text{dist}(f,\proPt\setminus\proP)\} \\
&\geq \min\{\eps,\eps_{0}\} \\
&\geq \eps_{0}/2.
\end{align*}
 
\item $f\in\proPt\setminus\proP$. Analogous to the case above, we have 
$\dist(f,\proP_1) \geq \eps_0/2$.
\item $f\notin\proPo\cup\proPt$. Then by triangle inequality,
$\max\{\text{dist}(f,\proPo\setminus\proP),\dist(f,\proPt\setminus\proP)\}\geq\eps_{0}/2$.
So there is some $i\in\{1,2\}$ such that $\dist(f,\proP_i\setminus\proP)\geq\eps_{0}/2$. 
Since $\dist(f,\proP)\geq \eps$, it follows that 
$\dist(f,\proP_i)\geq\min\{\eps,\eps_{0}/2\}=\eps_{0}/2$. 
\end{enumerate}
Thus, we conclude that there is some $i\in\{1,2\}$ such that $\dist(f,\proP_i)\geq \eps_0/2$.
This implies that $T'_i$, which makes at least $Q'_{i}(\eps)\geq q_{i}(\eps_{0}/2)$ queries, 
accepts $f$ with probability at most $1/3$. 
Hence, $T$ accepts $f$ with probability at most $1/3$ as well, completing the proof. \qed

\section{Proof of Proposition~\ref{prop:complement}}
\label{Sec:complement proof} 

We shall define a property $\proP=\{P_{2k}\}_{k>0}$, 
where $\proP_{2k} \subseteq \cube^{\F_2^{2k}}$ is
a collection of Boolean functions defined over $\F_2^{2k}$, 
such that neither $P_{2k}$ nor $\overline{P_{2k}}$ is testable.
Recall that a property $\proP$ is said to be \emph{testable} if there
is a tester for $\proP$ whose query complexity is independent of
the sizes of the inputs to the functions (in our case, 
independent of $k$).

First, let the Hadamard encoding $\Had:\kcube \times \kcube \rightarrow \cube$ be
$\Had(\alpha,x)=\alpha\cdot x$.
Note that $\F_{2^k}$ is isomorphic to $\F_2^k$, 
so for every function $g:\F_{2^k} \rightarrow \F_{2^k}$, 
the Hadamard concatenation of $g$ can be written as $\Had \circ g:\F_2^{2k}\rightarrow \cube$
where $(\Had\circ g)(x,y) \eqdef \Had(g(x),y)$.

We now define $P_{2k}$ as follows. Let $f \in P_{2k}$ 
if there exists a polynomial $p:\F_{2^k}\rightarrow \F_{2^k}$ 
of degree at most $2^{k-1}-1$ such that $\dist(f,\Had\circ p) < 1/8$. 
An important fact is that 
if $g:\F_{2^k}\rightarrow \F_{2^k}$ is a polynomial of degree $2^{k-1}$, 
then $\Had \circ g$ is not in $P_{2k}$. 
To see this, note that by the Schwartz-Zippel Lemma, 
if $q:\F_{2^k}\rightarrow \F_{2^k}$ is a polynomial of degree 
at most $2^{k-1}$, then  $\Pr_x[q(x)=0]\leq 1/2$. 
Therefore, for any polynomial $p$ of degree at most $2^{k-1}-1$, 
$\dist(p,g)\geq 1/2$.
This implies that $\dist(\Had\circ p, \Had \circ g) \geq 1/4$, 
since the Hadamard encoding has relative distance $1/2$.%
\footnote{In other words, suppose $x \in \F_{2^k}$ satisfies that 
$p(x)\neq g(x)$. Then the number of $y$'s such that 
$\Had(p(x),y)\neq \Had(g(x),y)$ is exactly $2^{k-1}$.}
Since the Hadamard encoding of $g$ is at least $1/4$-far 
from the Hadamard encoding of any degree $2^{k-1}-1$ polynomials, 
by construction of $P_{2k}$, $\Had \circ g$ is at least $1/8$-far from $P$, 
i.e., $\Had \circ g\in \overline{P_{2k}}$.

Now we show that neither $P_{2k}$ nor its complement is testable for 
any distance parameter $\eps<1/8$. 
By polynomial interpolation, for every set of $2^{k-1}-1$ points, 
there exists a polynomial of degree $2^{k-1}-1$ that agrees with $g$ on these points.
So any tester that distinguishes between members of $P_{2k}$ 
and members at least $\eps$-far away from $P_{2k}$ needs at least $2^{k-1}-1$ queries. 
Similarly, as we have just shown that $\Had \circ g\in \overline{P_{2k}}$
when $g$ is a degree-$2^{k-1}$ polynomial,
it follows that any tester that distinguishes between members of $\overline{P_{2k}}$ 
and functions at least $\eps$-far away from $\overline{P_{2k}}$ 
also need at least $2^{k-1}$ queries. 
To conclude, we have shown a property $\proP$
defined over domains of sizes $|\mathcal{D}|=2^{2k}$
but testing $\proP$ and $\overline{\proP}$
both require $\Omega(2^k)=\Omega(|\mathcal{D}|^{1/2})$ queries.
Thus, neither the property or its complement 
is testable with a query complexity independent of the sizes of the domains, 
completing the proof. 
\qed


\section{Proof of Proposition~\ref{thm:set-difference}}

\label{Sec:set-difference} 
Let $T_{1}$ be the tester for $\proPo$ with query complexity $q_{1}(\eps)$ 
and let $T_{2}$ be the tolerant tester for $\proPt$ with query complexity $q_{2}(\eps_1,\eps_2)$. 
First we convert $T_{1}$ into another tester $T'_{1}$ such that, 
on input distance parameter $\eps$, $T'_{1}$ makes $Q'_{1}(\eps)$ queries, where
\[
Q_1'(x)=\begin{cases}
q_1(x), & \text{ if $x<\eps_1$;}\\
\max\{q_1(x),q_i(\eps_1)\}, & \text{ otherwise.}
\end{cases}\]

Set $P=\proP_1\setminus \proP_2$ and define its tester $T$ as follows:
on input function $f$, $T$ first simulates $T_{1}$ and then $T_{2}$. 
$T$ accepts iff $T_1$ accepts and $T_2$ rejects. 
Since $\eps_1$ is a constant, $Q'_1(\eps))=O(\eps)$, and $T$ has query complexity $O(q_1+q_2)$.

For completeness, if $f\in\proP$, then by assumption $f\in\proPo$
and $\dist(f,\proPt)\geq\eps_{0}>\eps_2$. 
This implies that $T_1$ always rejects $f$, $T_2$ accepts $f$ with
probability at most $1/3$, and thus by a union bound argument $T$ accepts $f$ with probability at least $2/3$.

For soundness, suppose $\dist(f,\proP)\geq\eps$. 
We consider two cases and note that in both of them, $T$ accepts $f$ with probability at most $1/3$.

\paragraph{Case 1.} $\dist(f,\proP_2)\leq \eps_1$.

Since $T_2$ is a tolerant tester, $T_2$ rejects $f$ with probability at most $1/3$.
Thus, $T$ accepts with probability at most $1/3$ as well.
 
\paragraph{Case 2.} $\dist(f,\proP_2) > \eps_1$.

Since $\proP_1$ is the union of $\proP$ and $\proP_2$, we can conclude that 
$\dist(f,\proP_1) = \min\{\dist(f,\proP),\dist(f,\proP_2)\}$, which is at least 
$\min\{\eps,\eps_1\}$. 
Since $T'_1$ makes at least $\max\{q_1(\eps),q_1(\eps_1)\}$ queries, we know that
$T'_1$ accepts $f$ with probability at most $1/3$, and
hence, $T$ accepts $f$ with probability at most $1/3$ as well.
\qed
\end{document}